# Multi-microservice migration modelling, comparison, and potential in 5G/6G mobile edge computing: A non-average parameter values approach

Arshin Rezazadeh and Hanan Lutfiyya
Computer Science Department
Western University
London, Ontario, Canada
arezaza6@uwo.ca, hlutfiyy@uwo.ca

## Abstract

Integration of cloud, fog, and edge computing with future mobile Internet-of-Things (IoT) devices and related applications in 5G/6G networks will become more practical in the coming years. Virtual Memory (VM) was once a key approach in these contexts, but containers are quickly becoming the de facto virtualization technique in cloud-fog-edge services. Mobile IoT applications, e.g., intelligent transportation, virtual reality, healthcare, augmented reality, and online gaming, incorporating fog-edge, have increased the demand for a millisecond-scale response and processing time.

Multi-access Edge Computing (MEC) reduces remote network traffic while providing low latency for client-server communication. These services must run on MEC nodes that are physically close to devices in order to consistently provide such low latencies. However, classical migration techniques may not meet the requirements of future mission-critical IoT applications, which demand mobility support and a real-time response. IoT mobile devices have limited resources for running multiple containerized services, and client-server latency worsens when fog-edge services must migrate to maintain proximity in light of device mobility.

This study analyzes the performance of the MiGrror migration method and the pre-copy live migration method when the migration of multiple VMs/containers is considered. This paper presents mathematical models for the stated methods and provides migration guidelines and comparisons for services to be implemented as multiple containers, as in microservice-based environments. Experiments demonstrate that MiGrror outperforms the pre-copy technique and, unlike conventional live migrations, can maintain less than 10 milliseconds of downtime and reduce migration time with a minimal bandwidth overhead. The results show that MiGrror can improve service continuity and availability for users.

Most significant is that the model can use both average and non-average values for bandwidth, memory dirtying rate, and VM/container size during migration to achieve improved and more accurate results, while other research typically only uses average values. This paper shows that using only average parameter values in migration can lead to inaccurate results.

# 1 Introduction

New cloud-based applications (apps) have emerged due to the cloud's virtually infinite accessible resources and extensive service offerings [1–3]. Moreover, **microservices** are gaining increasing interest as a potent architectural practice for delivering software services. In this approach, applications are designed as a set of modules known as microservices, with each module focused on one component of the entire application [4]. Currently, microservices are delivered utilizing container frameworks rather than virtual machines (VMs). Although the microservices concept was originally built for the cloud context, it is gaining traction as a viable solution for **edge computing** environments [5]. However, these advancements have been followed by challenges for delay-sensitive applications with strict delay requirements [6]. **Mobility support** and **low latency** cannot be accommodated by the present cloud computing paradigm [7]. In order to solve these issues, the fog [8] and edge computing [9] paradigms have been proposed that seek to expand cloud resources and services and bring them closer to the network's edge where data is generated. Consequently, end-to-end latency is lower since the data is transmitted across fewer hops.

Multi-access/Mobile edge computing (MEC) was recently introduced as a key enabler of future 5G and 6G networks, shifting services from large remote cloud servers to an ubiquitous architecture of micro servers close to access networks and base stations [10–12]. This proximity can help MEC provide its main characteristics: **mobility support**, **real-time response**, and **high bandwidth** [13], which is especially important for mobile Internet-of-Things (IoT) devices. These characteristics are vital for demanding applications such as autonomous vehicles, healthcare, virtual reality, augmented reality, and online gaming [9,14–16], particularly when migration is involved. With more users shifting to edge computing and microservices, managing resources is becoming more challenging. Mobile devices at the network's edge may be repositioned between various MEC nodes. When this movement occurs, corresponding microservices may **require migration** between MEC nodes to keep proximity to the device [17]. Furthermore, some MEC nodes may become overloaded due to changing workloads, while others may stay underutilized on the same network infrastructure [18].

In some modern applications, multiple cooperating services are required in microservices-based environments to provide certain services; thus, we may need to consider migrating multiple microservices [19] for those containers that need to be in proximity to the device. Each containerized application may make use of multiple containers. Furthermore, each mobile IoT device may run multiple applications; in this environment, **multiple** container migration is inevitable [20,21]. Therefore, we need to investigate the simultaneous migration of multiple containers. Assume a smart city in which tourists are traversing with their mobile IoT devices. The mobile IoT devices are running applications such as augmented reality (AR) and virtual reality (VR) for a virtual tour guide in the context of the metaverse. The mobile IoT devices are connected to the edge to reduce application latency (turn-around time) and to provide more bandwidth. As a basic example of a containerized metaverse application, one microservice captures the environment from the device, and another microservice renders the AR data to the device. Each microservice can use single or multiple containers in its tasks. For the VR component of the application, another service deploys virtual reality components to the mobile device. This application needs ultra-low response time for smooth functionality. For such real-time applications, [22] suggests an end-to-end response threshold of 17 milliseconds ($ms$); otherwise, it cannot meet real-time latency requirements. While tourists traverse the city, mobile IoT devices require migration to keep their connections alive. Since the applications require high bandwidth and ultra-low latency, migrations and hand-offs must occur fast enough to keep the applications' response time as minimal as possible.

The hand-off is a migration component [15] that is triggered when a device disconnects from the access point (AP) of an edge node and connects to another node's AP on the same network infrastructure. **Downtime** occurs when a VM or container is unavailable during migration while a device is handed off from one edge node to the next [23]. Downtime caused by VM/container migrations lies in the range of seconds to minutes [15,24–30]. The **delay** is strongly affected by the amount of downtime and page faults [31,32]. Moreover, since the mobile IoT device must migrate from the old connection point to the new one throughout the procedure, it cannot access services or data during hand-off. There has been considerable work focused on reducing downtime [15,25–29,33]. Live migration techniques could facilitate downtime issues by sending and receiving data while the VM or container is still operating at the source or destination.

The pre-copy live migration proposed by Clark et al. [34], mostly used in literature, moves data from the source to the destination in pre-determined rounds that regularly transfer changes from the source to the destination. Despite pre-copy lowering the downtime compared to the non-live migration method, the VM/containers are not synchronized

(sync) immediately following a change in the memory from the source [23]. This late synchronization causes more data to be required to transfer after hand-off and, consequently, high downtime and migration time for delay-sensitive applications [23]. Some or all application components of intelligent transportation, virtual reality, healthcare, augmented reality, and online gaming requires ultra-low latency for data processing and communication [35].

The stated end-to-end response is difficult to achieve with the pre-copy method which led Rezazadeh et al. [23] proposed **MiGrror migration technique** for faster synchronization between source and destination, which results in less data transmitted during hand-off and, consequently, less migration time and downtime compared to the pre-copy method. The MiGrror technique mirrors memory from the source to the destination in the same way that mirroring is used in wide-area network servers. In this analysis, the pre-copy method ends and hands off when the number of rounds reaches a pre-defined threshold, e.g., 10-30 rounds in most research. To ensure fairness, we initiate the hand-off for both methods at the same time: pre-copy and MiGrror.

Furthermore, another limitation of the previous work on migration is that the evaluation typically assumes average input parameter values during migrations. The input parameters we take into account in this research are the **transfer rate** (provisioned bandwidth), **memory dirtying rate** and memory **size** of the VM/container (VM/container size), which, considering **mobility**, can constantly change during migration for each VM/container of an application. However, the given input parameter values **fluctuate** over time. Memory dirtying rate and container size values vary during the migration process depending on the task for each container. Moreover, the transfer rate can vary throughout the migration since the user's mobility causes changes in distance between the user's device and its services, resulting in diverse signal strength and available bandwidth [36].

Studies [18,19,33,37–51] use **migration modelling** to comprehend the future behaviour of a system. Although most research employs average parameter values [18,19,33,37–51] and **assumes** the **input parameter values remain unchanging**, our study demonstrates that the results vary since the input parameters can constantly change during the migration. Our results show it is essential to learn if these parameters are higher or lower at the beginning, middle, and end of the migration, considering migration time and downtime as output parameters. Downtime and migration time are the two primary output parameters in migration [18].

As a result, using only average parameter values can result in output parameters that differ from their realistic outputs because output parameters can be similar, when using the same average values for input parameters. In contrast, the outputs for non-average input parameters **can deviate**, while maintaining the same average input values. Consequently, utilizing non-average parameter values can produce different results while their averages remain unchanged. Using non-average input parameter values can result in more precise migration time and downtime outcomes. In addition, innovative strategies for migrating multiple VMs/containers are possible when considering non-average parameter values.

To exemplify the discussed **current migration models' limitations**, consider two migration procedures with the same average parameter values (e.g., transfer rate and memory dirtying rate) but varying values during the migration process. In this example, increasing the memory dirtying rate at the end of the migration significantly impacts downtime since downtime occurs when the migration is complete [17], and this increased memory dirtying rate requires transferring a higher-than-average memory dirtying rate. The same holds true when we decrease the transfer rate at the end of the migration and the dirtied memory data transfers at a lower-than-average transfer rate. In both cases, using average parameter values generates the same migration time and downtime, whereas using actual (non-average) parameter values generates different outputs.

The same concern as described above occurs due to employing average parameter values in current migration models while we decrease the actual (non-average) transfer rate at the beginning of the migration in this example. As a result, if input parameters are higher or lower than their average value at crucial points - at the beginning and the end of the migration process, they can negatively affect downtime and migration time. These situations worsen when parameters at the given critical migration points are significantly higher or lower than average, and in these cases, they extremely affect downtime and migration time.

However, **these crucial states are hidden** when developing the edge computing environment relying exclusively on average input parameter values. As a result, **new migration models in edge computing are required** to achieve more accurate results for multi-container mission-critical 5G/6G mobile IoT applications. This paper presents mathematical models for the pre-copy and MiGrror migration methods, offering a new perspective on modelling by utilizing both average and non-average values during MiGrror migration, while typically, only average values are considered in the literature [18,19,33,37–51]. This research takes into account non-average values of transfer rate, memory dirtying rate and VM/container size for the MiGrror technique in addition to average values. Our experiments show that the results of both average and non-average parameter values for the pre-copy method are mostly identical since memory changes several times in each round. However, the MiGrror method can take into account a larger number of synchronizing events, which is advantageous since using actual (non-average) parameter values rather than average ones is possible when employing the MiGrror method. We take this novel approach since some parameters, such as memory dirtying rate, may change several times during the migration. To the best of our knowledge, this is the first time that different values of bandwidth, memory dirtying rate, and VM/container size, rather than classical average values, are considered during the migration of each single VM or container. To distinguish between these two types of modelling, we also consider average parameter value results and compare them to non-average parameter value results in section 6. Furthermore, the non-average MiGrror migration model is applicable regardless of whether machine learning approaches, compression, or other methods are used to decrease migration time and downtime.

In this paper, we first model the migration of multiple containers for both stated migration methods. We do this by first using average values of the CSAP dataset [52], then non-average values, followed by non-dataset input values. We also compare the migration overhead of both methods listed and discuss which is better suited to specific scenarios.

The main contributions of this study are summarized as follows:

- We present the MiGrror mathematical migration **model** for heterogeneous **multiple** VMs/containers. This is the first MiGrror model that considers the simultaneous migration of multiple VMs/containers.
- For the first time, we use **non-average** and classical average values for the transfer rate, memory dirtying rate, and VM/container size, during each migration period of every single VM/container for the MiGrror method.
- We conducted experiments to analyze the input parameters that impact the performance of the investigated migration methods.

The remainder of this paper is structured as follows: Section 2 delivers background and related work on classic migration strategies. Section 3 describes the classic pre-copy live migration model used in this analysis to compare with the new model. Section 4 presents models of the MiGrror migration for multiple VMs/containers. Section 5 provides evaluations and discussions, and Section 6 concludes this study.

# 2 Background and related work

This section provides a high-level overview of edge computing live migration techniques, as well as models for migrating multiple VMs and containers. Live migration allows virtual machines and containers to remain operational for most of the migration process [39]. First the pre-copy and post-copy live migration methods are summarised, followed by the MiGrror technique, and finally, the migration model studies are reviewed.

**Pre-copy live migration technique:** With **pre-copy migration** [34], the entire VM/container state is sent from the current node to the target node. An **iteration** is a round in which the pre-copy waits for memory changes to send at the end of each round. The source then resends **dirty pages**, which are updated memory pages from the previous iteration, over a number of iterations. Upon receiving the hand-off signal, the source VM/container pauses execution to prevent memory and state modification and transfers the final dirty page and the latest changes in the runtime (execution) state, which includes CPU and register updates, to the target edge node. Finally, the VM/container resumes operation on the target edge node. Since the pre-copy technique typically transmits each memory page multiple times, it may have a negative impact on the **total amount of data transmitted** throughout the migration process and, consequently, the **total migration time** [23]. Figure 1 shows the pre-copy iterations and related symbols.

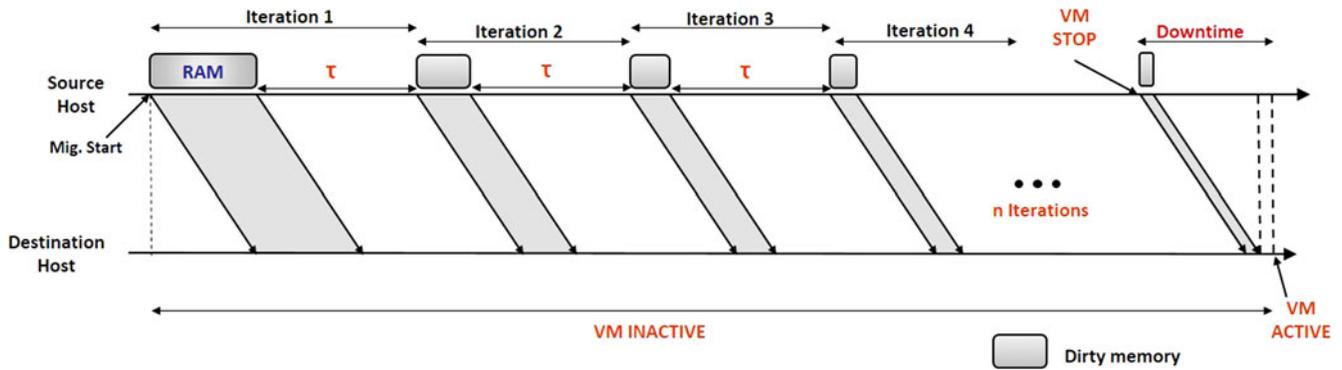

Fig 1.  Pre-copy iterations (rounds) [51]

**MiGrror migration technique:** The **MiGrror** migration method was introduced in [23] to reduce migration time and downtime when compared to the pre-copy method. This objective is accomplished by synchronizing the source and destination more frequently, resulting in a **mirror** of the VM/container at the destination, similar to how mirroring is done in wide-area network servers [23]. This technique reduces the amount of data transferred during hand-off. Despite the intention to use more bandwidth, the results indicate that this method outperforms pre-copy in terms of downtime, delay, and total migration time. Furthermore, the amount of data transferred during migration is greater than that of live migration techniques. Since 5G and 6G networks have significantly more available bandwidth than previous generations, increasing bandwidth usage between MEC nodes in this approach should not substantially impact overall performance [23,53]. This method will be examined by presenting a mathematical model in section 4, followed by results and discussion in section 6.

**Post-copy live migration technique:** Before transferring the latest state from the source to the target, the **post-copy migration** technique [54] pauses the VM/container execution to prevent runtime state changes. The state is then transferred to the target, along with the minimum memory and state required to resume the execution of the VM/container. The VM/container is then resumed at the target. An access problem occurs when the VM attempts to access a page that the target has not yet received. In this condition, a page fault occurs, and the source transmits the faulty page to the target. When the VM/container is restarted at the target node during the post-copy process, any applications executing in the VM/container continue to run at the target. After sending all remaining pages, the page transfers to the target stops, and the VM/container post-copy migration is complete.

**Modelling live migration techniques:** Several research studies on live migration modelling have been conducted over the last decade. Most of them base their research on a **single** VM or container migration [33,46–51]. They primarily focus on downtime and migration time and compare live migration techniques based on various input values, such as pre-copy iterations, page dirtying rate, bandwidth, and VM/container size, using datasets, implementations, or their assumptions. A subset of these papers provides models and compares various parameters of live migration methods [46–49], while others employ estimation and optimization techniques to reduce migration costs, such as downtime and migration time [33,50,51]. Despite extensive research on modelling the migration of a single VM/container, few authors focus on modelling multiple-VM/multiple-container migration [18,19,37–45]. Some of these studies focus on the number of VMs/containers and provisioned bandwidth in addition to the stated input values. Most of these studies focus on modelling **multiple-VM** migration [18,19,37,38,40,42–45] in order to optimize the migration performance of multiple VMs, while authors in [39,41] focus on modelling **multiple-container** migration. The authors in [18,19,33,37–51] employ only average parameter values, whereas we use non-average values for more precise results.

**Single service migration:** Altahat et al. [33] propose a neural network-based model that predicts VM migration performance metrics for pre-copy and post-copy methods as well as different application workloads to analyze the migration models under various workloads. Metrics include downtime, migration time, and the amount of data transferred during the migration process. They compare their model to Linear Regression, SVR, and SVR with bagging. The authors of [48] propose an adaptive VM monitoring strategy for migrating a single VM using pre-copy and post-copy methods. They develop an autoregressive model to predict the dirty memory rate and use it to reduce migration downtime, migration time, and the data transfer amount. The model's output value is determined by a linear combination of a stochastic variable and the previous model's values. Tang et al. [49] use reinforcement learning with deep Q-learning container migration to propose power consumption, delay, and migration cost models. They compare their algorithm to other ML algorithms, including static threshold, median absolute deviation, and interquartile range regression. Baccarelli et al. [50] use the pre-copy migration time, downtime, round-trip time, and energy consumption models to reduce delay and energy consumption in wireless connections with a bandwidth manager.

**Multiple service migration:** The authors develop adaptive bandwidth allocation in [18,39,40] to minimize migration time in their models. Singh et al. [18] use Geometric Programming to assign transfer and compression rates to each VM in order to reduce the total migration time of multi VMs. They consider parameters including VM size, memory dirtying rate, transfer rate, and compression ratio of VMs. They evaluate their experiments with up to seven VMs and nine pre-copy iterations. Maheshwari et al. [39] developed a cost model for multi-container migration, considering container size, number of containers, memory dirtying rate, bandwidth, and load at an edge node that supports mobility. They use a Min-Max model to minimize the migration cost. Liu et al. [40] build a migration cost model by predicting the memory dirtying rate and employing parameters such as VM size and transfer rate. They use a cost model for multi-VM and employ adaptive bandwidth allocation to reduce migration costs.

The following papers consider migration to be the primary or secondary contribution of their analysis. In more detail, Sun et al. [19] use an M/M/C/C queuing migration model to optimize migration and reduce downtime and migration time for multiple VMs. Satpathy et al. [37] compare migration model performance for multiple VMs, including comparisons based on VM size, memory dirtying rate, and available bandwidth. Using a platform based on a software-defined network (SDN), He et al. [38] evaluate the performance of multiple VM migration models. They consider migration time and downtime to be two of their most important criteria. To balance server load, Zhang et al. [41] propose a set of algorithms for optimizing load balancing and migrating multiple containers among cloud servers in order to balance server load. Their primary focus is load balancing; migration would occur as a result of server load balancing with the migration time model. Similarly, Forsman et al. [44] present a load-balancing solution that reduces the migration cost of multiple VMs. They also include migration time and downtime in their cost model. In another study, Satpathy et al. [42] propose a VM placement strategy for cloud servers while modelling multiple-VM migration with downtime and migration time. Considering power constraints, Elsaid et al. [43] examine the migration cost of multiple VMs using migration time and power consumption. Cerroni [45] investigates the cost of migrating multiple VMs based on downtime and migration time using the Markovian model. The network overhead and throughput degradation are also components of the migration cost model.

# 3 A pre-copy migration mathematical model for multiple VMs/containers using average parameter values - the pre-copy migration model

This section describes existing pre-copy migration models that use average parameter values, similar to [18,19,33,37–40,46–51]. We use only average parameter values for the pre-copy method since considering non-average parameter values is ineffective. This method typically employs a limited number of rounds, e.g. 10-30 rounds in most research; however, this is incompatible with constantly changing parameter values, such as memory dirtying and transfer rates. The transfer rate and memory can change multiple times in each round of the pre-copy migration in short intervals. Therefore, there is no single value for the stated parameters for each round of the pre-copy. Furthermore, since the pre-copy migration method uses a limited number of rounds, results for both average and non-average parameter values are mostly identical. Utilizing non-average parameter values based on the preceding discussions would be inefficient, as the pre-copy method cannot employ every single value of each input parameter during the migration. Although the stated pre-copy migration models are not completely identical, the pre-copy migration for multiple VMs/containers in this section will be modelled derived from the models used in [18,19,33,37–40,46–51], so that we can compare the pre-copy results with the proposed migration model. Downtime and migration time are the two primary parameters for migration modelling analysis [18]. Downtime is an important performance metric for end-users, which must be as low as possible to avoid service interruptions [18]. The total migration time must be as short as possible because it consumes computational and network resources from both the origin and the target MEC nodes [18]. The amount of data that must be transmitted during the migration of multiple VMs/containers is also considered as an overhead metric of the migration process in this paper. Table 1 defines a number of key parameters and their notations for the models described in this paper. In the table, $M_j$, $\bar{d}_j$, and $\bar{r}_j$ represent the VM/container memory size, average memory dirtying rate, and average transfer rate (average bandwidth) available during migration for any $VM_j/container_j$, respectively. The parameters specified affect migration time $\left(TM_j^{Pre}\right)$ and downtime $\left(TD_j^{Pre}\right)$. Higher

$M_j$ and $\bar{d}_j$ levels increase migration time and downtime, while higher $\bar{r}_j$ levels decrease migration time and downtime. The remainder of this section will examine the **downtime**, **migration time**, and **migration overhead** of the pre-copy migration model.

### Table 1

Symbols and Definitions

| Parameter | Description |
|---|---|
| $m$ | The number of migration iterations in the pre-copy method |
| $n$ | The number of migration events in the MiGrror method |
| $p$ | The number of VMs/Containers to be migrated |
| $V_{i,j}^{Pre}$ | Dirty memory generated during round $i$ for any $VM_j/Container_j$ in pre-copy, $\forall j \in \{1, ..., p\}$ and $\forall i \in \{1, ..., m\}$ |
| $V_{i,j}^{Mir}$ | Dirty memory generated during event $i$ for any $VM_j/Container_j$ in MiGrror, $\forall j \in \{1, ..., p\}$ and $\forall i \in \{1, ..., n\}$ |
| $V_{s,j}^{Pre}, V_{s,j}^{Mir}$ | Memory amount in the stop-and-copy phase for any $VM_j/Container_j$ in pre-copy and MiGrror migration methods |
| $M_j$ | Memory size of any $VM_j/Container_j$, $\forall j \in \{1, ..., p\}$ |
| $\bar{r}_j$ | Average transfer rate (average bandwidth) available for any $VM_j/Container_j$ in pre-copy migration method |
| $r_{i,j}$ | Available transfer rate (bandwidth) during event $i$ for any $VM_j/Container_j$ in MiGrror migration method |
| $r_{s,j}$ | Available transfer rate (bandwidth) in the stop-and-copy phase for any $VM_j/Container_j$ in MiGrror migration method |
| $t_{i,j}^{Pre}$ | Time to transfer $V_{i,j}^{Pre}$ in pre-copy migration, $\forall j \in \{1, ..., p\}$ and $\forall i \in \{1, ..., m\}$ |
| $t_{i,j}^{Mir}$ | Time to transfer $V_{i,j}^{Mir}$ in MiGrror migration, $\forall j \in \{1, ..., p\}$ and $\forall i \in \{1, ..., n\}$ |
| $\bar{d}_j$ | Average memory dirtying rate for any $VM_j/Container_j$ in pre-copy migration method |
| $d_{i,j}$ | Memory dirtying rate during event $i$ for any $VM_j/Container_j$ in MiGrror migration method |
| $TM_j^{Pre}, TM_j^{Mir}$ | Total migration time for any $VM_j/Container_j$ for both pre-copy and MiGrror methods |
| $TD_j^{Pre}, TD_j^{Mir}$ | Downtime for any $VM_j/Container_j$ for both pre-copy and MiGrror migration methods |
| $\lambda_j$ | $\bar{d}_j/\bar{r}_j$, $\forall j \in \{1, ..., p\}$ |
| $\lambda_{i,j}$ | $d_{i-1,j}/r_{i,j}$, $\forall j \in \{1, ..., p\}$ and $\forall i \in \{1, ..., m\}$ |
| $\lambda_{s,j}$ | $d_{n,j}/r_{s,j}$ |
| $B$ | Total maximum bandwidth reserved for the entire migration between two MEC nodes |
| $TA_j^{Pre}, TA_j^{Mir}$ | Migration overhead, amount of data to be migrated during migration, for any $VM_j/Container_j$ in pre-copy and MiGrror methods |
| $\tau$ | The inter-iteration delay in pre-copy migration method |
| $\tau_{i,j}$ | The time between two consecutive events $i$ and $i+1$ in MiGrror migration method for $VM_j/Container_j$ |

During round one, the entire memory of any $VM_j$ or $container_j$ is transferred from the source to the destination. As a result, the data transmitted during round one, i.e., $V_{1,j}^{Pre}$, may be calculated using the equation below:

$$V_{1,j}^{Pre} = M_j \tag{1}$$

The memory becomes dirty throughout the transfer as the $VM/container$ remains active at the source during pre-copy migration. Then, the pre-copy rounds transfer just the memory that was dirtied during the preceding round. The amount of data sent at round $i$ for every $VM_j/container_j$ is:

$$V_{i,j}^{Pre} = \bar{d}_j t_{i-1,j}^{Pre} \tag{2}$$

As soon as $i$ reaches $m$, the final round, i.e. stop-and-copy, begins. We assume that all VMs and containers have $m$ rounds and that every single $VM_j/container_j$ stops execution after $m$ rounds before the stop-and-copy phase. Furthermore, the time necessary for the transfer round $i$ for $VM_j/container_j$, i.e., $t_{1,j}^{Pre}$, may be recursively calculated using equations (1) and (2) as follows:

$$t_{1,j}^{Pre} = \frac{V_{1,j}^{Pre}}{\bar{r}_j} + \tau = \frac{M_j}{\bar{r}_j} + \tau \tag{3}$$

where $\tau$ is the inter-iteration delay shown in Figure 2.

$$t_{2,j}^{Pre} = \frac{V_{2,j}^{Pre}}{\bar{r}_j} + \tau = \frac{\bar{d}_j t_{1,j}^{Pre}}{\bar{r}_j} + \tau = \lambda_j t_{1,j}^{Pre} + \tau = \lambda_j \left(\frac{M_j}{\bar{r}_j} + \tau\right) + \tau = \frac{\lambda_j M_j}{\bar{r}_j} + \lambda_j \tau + \tau = \frac{\lambda_j M_j}{\bar{r}_j} + \tau(1 + \lambda_j) \tag{4}$$

where $\lambda_j$ is the average memory dirtying rate divided by the average transfer rate, $\bar{d}_j/\bar{r}_j$, for any $VM_j/container_j$.

$$t_{3,j}^{Pre} = \frac{V_{3,j}^{Pre}}{\bar{r}_j} + \tau = \frac{\bar{d}_j t_{2,j}^{Pre}}{\bar{r}_j} + \tau = \lambda_j t_{2,j}^{Pre} + \tau = \frac{M_j}{\bar{r}_j}\lambda_j^2 + \tau\left(\frac{1 - \lambda_j^3}{1 - \lambda_j}\right) \tag{5}$$

...

$$t_{i,j}^{Pre} = \frac{V_{i,j}^{Pre}}{\bar{r}_j} + \tau = \lambda_j t_{i-1,j}^{Pre} + \tau = \frac{M_j}{\bar{r}_j} \lambda_j^{i-1} + \tau \left( \frac{1 - \lambda_j^i}{1 - \lambda_j} \right) \qquad (6)$$

Thus, the migration **downtime** for $VM_j/container_j$, i.e., $TD_j^{Pre}$, may be calculated as:

$$TD_j^{Pre} = \frac{V_{s,j}^{Pre}}{\bar{r}_j} = \frac{\bar{d}_j t_{m,j}^{Pre}}{\bar{r}_j} = \lambda_j t_{m,j}^{Pre} = \frac{M_j}{\bar{r}_j} \lambda_j^m + \lambda_j \tau \left( \frac{1 - \lambda_j^m}{1 - \lambda_j} \right) \qquad (7)$$

where $V_{s,j}^{Pre}$ represents the data during hand-off for any $VM_j/container_j$. We use a maximum value here since containers are dependent on, and interact with, one another, and some must wait for others to respond to each user. The maximum amount of downtime during pre-copy migration is expressed as follows:

$$T_{Downtime}^{Pre} = max\{TD_1^{Pre}, TD_2^{Pre}, TD_3^{Pre}, \ldots, TD_p^{Pre}\} \qquad (8)$$

Further, the total **migration time** for every $VM_j/Container_j$, i.e., $TM_j^{Pre}$ with $m$ number of pre-copy transfer rounds followed by a final stop-and-copy round, is given by:

$$\begin{aligned}
TM_j^{Pre} &= \sum_{i=1}^{m} t_{i,j}^{Pre} + TD_j^{Pre} = \left( \frac{M_j}{\bar{r}_j} \sum_{i=1}^{m} (\lambda_j)^{i-1} + \frac{\tau}{1 - \lambda_j} \sum_{i=1}^{m} (1 - \lambda_j)^i \right) + TD_j^{Pre} \\
&= \frac{M_j}{\bar{r}_j} \frac{1 - \lambda_j^m}{1 - \lambda_j} + \tau \frac{m(1 - \lambda_j) - \lambda_j (1 - \lambda_j^{m+1})}{(1 - \lambda_j)^2} + TD_j^{Pre} \\
&= \frac{M_j}{\bar{r}_j} \frac{1 - \lambda_j^m}{1 - \lambda_j} + \tau \frac{m(1 - \lambda_j) - \lambda_j (1 - \lambda_j^{m+1})}{(1 - \lambda_j)^2} + \frac{M_j}{\bar{r}_j} \lambda_j^m + \lambda_j \tau \left( \frac{1 - \lambda_j^m}{1 - \lambda_j} \right)
\end{aligned} \qquad (9)$$

The maximum migration time by assigning network transfer rate $\bar{r}_j$ for each $VM_j/container_j$ in pre-copy migration can be expressed as:

$$T_{migration}^{Pre} = max\{TM_1^{Pre}, TM_2^{Pre}, TM_3^{Pre}, \ldots, TM_p^{Pre}\} \qquad (10)$$

Thus, the total amount of data, **migration overhead**, to be sent during migration for any $VM_j/container_j$, i.e., $TA_j^{Pre}$, is given by:

$$\begin{aligned}
TA_j^{Pre} &= \sum_{i=1}^{m} V_{i,j}^{Pre} + V_{s,j}^{Pre} = M_j + \sum_{i=2}^{m} \bar{d}_j t_{i-1,j}^{Pre} + \bar{d}_j t_{m,j}^{Pre} = M_j + \left( \bar{d}_j \sum_{i=2}^{m} t_{i-1,j}^{Pre} \right) + \bar{d}_j t_{m,j}^{Pre} \\
&= M_j + M_j \lambda_j \frac{1 - \lambda_j^m}{1 - \lambda_j} + \tau \bar{d}_j \frac{m(1 - \lambda_j) - \lambda_j (1 - \lambda_j^{m+1})}{(1 - \lambda_j)^2} + \bar{d}_j t_{m,j}^{Pre} \\
&= M_j + M_j \lambda_j \frac{1 - \lambda_j^m}{1 - \lambda_j} + \tau \bar{d}_j \frac{m(1 - \lambda_j) - \lambda_j (1 - \lambda_j^{m+1})}{(1 - \lambda_j)^2} + M_j \lambda_j^m + \bar{d}_j \tau \left( \frac{1 - \lambda_j^m}{1 - \lambda_j} \right)
\end{aligned} \qquad (11)$$

The total migration overhead during migration for all VMs/containers is expressed as follows:

$$Data_{migration}^{Pre} = \{TA_1^{Pre} + TA_2^{Pre} + TA_3^{Pre} + \cdots + TA_p^{Pre}\} \qquad (12)$$

For downtime, migration time, and total migration overhead, subject to:

$$\sum_{j=1}^{p} \bar{r}_j \leq B \qquad (13)$$

where $B$ is the total maximum reserved bandwidth for the entire migration between two edge (MEC) nodes, and:

$$0 \leq \bar{r}_j \leq B \text{ and } \lambda_j < 1 \qquad (14)$$

# 4 Mathematical model of multiple VMs/containers migration using non-average parameter values – MiGrror model

This section describes the MiGrror [23] migration model, which for the first time, uses non-average parameter values for transfer rate (bandwidth), memory dirtying rate, and VM/container size during migration. Downtime, migration time, and migration overhead (the amount of data that must be transferred during migration) of multiple VMs/containers are all modelled. Since stated parameters, such as memory dirtying rate, are likely to change and do not have a fixed value during the migration process, the use of non-average parameter values can lead to more accurate migration time and downtime results. Although other researchers used average parameter values, our proposed migration time and downtime models revealed that the results would be different if these parameters were higher or lower at the beginning, middle, and end of the migration process with the same average parameter value. The problem with relying solely on average parameter values is that the migration and downtime results will be identical, while these results for non-average parameter values will vary. In contrast, if these parameters are higher or lower at the beginning of the migration, they can significantly affect migration time with the same average parameter values. They can also significantly impact downtime if they change at the end of the migration process with the same average parameter values. These findings imply that it is advantageous to know when the value of a parameter has a more significant impact on the result and that we can control the result by precisely selecting other parameter values, when it is possible, to achieve desired results. As a result, using non-average parameter values can provide greater insight and control over the migration process, particularly for 5G and 6G networks. Table 1 describes the modelling parameters. In the table, $M_j$, $d_{i,j}$, and $r_{i,j}$ represent the memory size, memory dirtying rate during event $i$, and available transfer rate during event $i$ in migration for any $VM_j/container_j$, respectively. The specified parameters affect migration time $(TM_j^{Mir})$ and downtime $(TD_j^{Mir})$. Higher $M_j$ and $d_{i,j}$ levels increase migration time and downtime, whereas higher $r_{i,j}$ levels decrease migration time and downtime. It is also critical that different levels of $r_{i,j}$ and $d_{i,j}$ occur at the beginning, end (during hand-off), as well as the middle of the migration process. The $r_{i,j}$ and $d_{i,j}$ levels are more critical for migration time at the beginning of the migration process, and these levels are more critical for downtime at the end of the migration process. The levels of both stated parameters have the least impact on migration time and downtime in the middle. The remainder of this section presents the MiGrror model.

During the first event, the entire memory of any $VM_j$ or $container_j$ is transferred from the source to the destination. So, the data sent during event one, i.e., $V_{1,j}$, can be computed using the equation given below:

$$V_{1,j}^{Mir} = M_j \tag{15}$$

The memory becomes dirty throughout the transfer as the $VM/container$ remains active at the source during MiGrror migration. Then, memory-change events transfer just the memory that was dirtied during the preceding event. The amount of data sent at event $i$ for every $VM_j/container_j$ is:

$$V_{i,j}^{Mir} = d_{i-1,j} t_{i-1,j}^{Mir} \tag{16}$$

Furthermore, the time necessary for the transfer event $i$ for $VM_j/container_j$, i.e., $t_{1,j}^{Pre}$, may be recursively calculated using equations (15) and (16) as follows:

$$t_{1,j}^{Mir} = \frac{V_{1,j}}{r_{1,j}} + \tau_{1,j} = \frac{M_j}{r_{1,j}} + \tau_{1,j} \tag{17}$$

where $r_{1,j}$ is the available transfer rate during the first event in migration for $VM_j/container_j$, and $\tau_{1,j}$ is the time between the first and second consecutive events of MiGrror migration for $VM_j/container_j$.

$$t_{2,j}^{Mir} = \frac{V_{2,j}^{Mir}}{r_{2,j}} + \tau_{2,j} = \frac{d_{1,j} t_{1,j}^{Mir}}{r_{2,j}} + \tau_{2,j} = \lambda_{2,j} t_{1,j}^{Mir} + \tau_{2,j} = \lambda_{2,j}\left(\frac{M_j}{r_{1,j}} + \tau_{1,j}\right) + \tau_{2,j} = \lambda_{2,j} \frac{M_j}{r_{1,j}} + \lambda_{2,j}\tau_{1,j} + \tau_{2,j} \tag{18}$$

$$t_{3,j}^{Mir} = \frac{V_{3,j}^{Mir}}{r_{3,j}} + \tau_{3,j} = \frac{d_{2,j} t_{2,j}^{Mir}}{r_{3,j}} + \tau_{3,j} = \lambda_{3,j} t_{2,j}^{Mir} + \tau_{3,j} = \lambda_{3,j}\lambda_{2,j}\frac{M_j}{r_{1,j}} + \lambda_{3,j}\lambda_{2,j}\tau_{1,j} + \lambda_{3,j}\tau_{2,j} + \tau_{3,j} \tag{19}$$

$$t_{4,j}^{Mir} = \frac{V_{4,j}^{Mir}}{r_{4,j}} + \tau_{4,j} = \lambda_{4,j} t_{3,j}^{Mir} + \tau_{4,j} = \lambda_{4,j}\lambda_{3,j}\lambda_{2,j}\frac{M_j}{r_{1,j}} + \lambda_{4,j}\lambda_{3,j}\lambda_{2,j}\tau_{1,j} + \lambda_{4,j}\lambda_{3,j}\tau_{2,j} + \lambda_{4,j}\tau_{3,j} + \tau_{4,j} \quad (20)$$

...

$$t_{i,j}^{Mir} = \frac{V_{i,j}^{Mir}}{r_{i,j}} + \tau_{i,j} = \lambda_{i,j} t_{i-1,j}^{Mir} + \tau_{i,j} \quad (21)$$

where $\tau_{i,j}$ is the time between two consecutive events $i$ and $i+1$ in MiGrror migration for $VM_j/container_j$. Moreover, $\lambda_{i,j}$ is the memory dirtying rate of the previous event, event $i-1$, divided by the transfer rate of the current event, event $i$, for any $VM_j/container_j$, and is equal to $d_{i-1,j}/r_{i,j}$. Then:

$$t_{i,j}^{Mir} = \lambda_{i,j}\lambda_{i-1,j}\ldots\lambda_{2,j}\frac{M_j}{r_{1,j}} + \lambda_{i,j}\lambda_{i-1,j}\ldots\lambda_{2,j}\tau_{1,j} + \lambda_{i,j}\lambda_{i-1,j}\ldots\lambda_{3,j}\tau_{2,j} + \lambda_{i,j}\lambda_{i-1,j}\ldots\lambda_{4,j}\tau_{3,j} + \cdots \\ + \lambda_{i,j}\lambda_{i-1,j}\tau_{i-2,j} + \lambda_{i,j}\tau_{i-1,j} + \tau_{i,j} \quad (22)$$

Thus, the migration **downtime** for $VM_j/container_j$ i.e., $TD_j^{Mir}$ can be calculated as:

$$TD_j^{Mir} = \frac{V_{s,j}^{Mir}}{r_{s,j}} = \frac{d_{n,j} t_{n,j}^{Mir}}{r_{s,j}} = \lambda_{s,j} t_{n,j}^{Mir} \\ = \lambda_{s,j}\left(\prod_{i=2}^{n} \lambda_{i,j}\frac{M_j}{r_{1,j}} + \prod_{i=2}^{n} \lambda_{i,j}\,\tau_{1,j} + \prod_{i=3}^{n} \lambda_{i,j}\,\tau_{2,j} + \prod_{i=4}^{n} \lambda_{i,j}\,\tau_{3,j} + \cdots + \lambda_{n,j}\lambda_{n-1,j}\tau_{n-2,j}\right. \\ \left. + \lambda_{n,j}\tau_{n-1,j} + \tau_{n,j}\right) \quad (23)$$

where $V_{s,j}^{Mir}$ and $r_{s,j}$ represent the data sent and the available transfer rate during hand-off, respectively, and $\lambda_{s,j}$ is $d_{n,j}/r_{s,j}$ for any $VM_j/container_j$. We use a maximum value here since containers are dependent on, and interact with, one another, and some must wait for others to respond to each user. The maximum amount of downtime during MiGrror migration is expressed as follows:

$$T_{Downtime}^{Mir} = max\{TD_1^{Mir},\ TD_2^{Mir}, TD_3^{Mir},\ \ldots,\ TD_p^{Mir}\} \quad (24)$$

Further, the total MiGrror **migration time** for every $VM_j/container_j$, i.e., $TM_j^{Mir}$ with $n$ number of transfer events followed by a final stop-and-copy event, is given by:

$$TM_j^{Mir} = \sum_{i=1}^{n} t_{i,j}^{Mir} + TD_j^{Mir} \quad (25)$$

The total migration time by assigning network transfer rate $r_{i,j}$ for each $VM_j/Container_j$ in MiGrror migration can be expressed as:

$$T_{migration}^{Mir} = max\{TM_1^{Mir},\ TM_2^{Mir}, TM_3^{Mir},\ \ldots,\ TM_p^{Mir}\} \quad (26)$$

Thus, the maximum amount of data, **migration overhead**, to be sent during MiGrror migration for any $VM_j/container_j$, i.e., $TA_j^{Mir}$, is given by:

$$TA_j^{Mir} = \sum_{i=1}^{n} V_{i,j}^{Mir} + V_{s,j}^{Mir} = M_j + \sum_{i=2}^{n} d_{i-1,j} t_{i-1,j}^{Mir} + d_{n,j} t_{n,j}^{Mir} \quad (27)$$

The total migration overhead during migration for all VMs/containers is expressed as follows:

$$Data_{migration}^{Mir} = \{TA_1^{Mir} + TA_2^{Mir} + TA_3^{Mir} + \cdots + TA_p^{Mir}\} \quad (28)$$

For downtime, migration time, and total migration overhead, subject to:

$$\sum_{j=1}^{p} r_{i,j} \leq B \quad (29)$$

where $B$ is the total maximum reserved bandwidth for the entire migration between two edge (MEC) nodes, and:

$$0 \leq r_{i,j} \leq B \text{ and } \lambda_{i,j} < 1 \tag{30}$$

## 5 Performance evaluation and discussions

Several parameters may affect migration performance, including container size, transfer rate, and memory dirtying rate. This section investigates how various parameters affect migration performance. We use the CSAP dataset [52] and our experiments to model pre-copy and MiGrror migration methods. The migration time, downtime, and migration overhead (transferred data) numbers given in the results are the averages of ten distinctive migration runs of each model using the Python code we developed. The pre-copy method terminates when the number of rounds (iterations) reaches a predefined threshold of 10 rounds ($m = 10$). To be fair in our comparisons, we trigger the hand-off for both migration methods at the same time. Furthermore, we use 20 VMs/containers ($p = 20$) to migrate from the source to the destination during migration. We divide the total bandwidth ($B$) by the number of VMs/containers ($p$) for non-dataset values, and the transfer rate for each $VM_j/container_j$ is the same. Additional considered parameters vary and are detailed in the subsections that follow.

Since we, unlike other researchers, consider non-average parameter values, we calculate the minimum, maximum, median, average, and standard deviations of the stated parameters to examine the dataset in more depth. The transfer rate $(r_{i,j})$ ranges between a minimum of 50 $megabits\ per\ seconds$ ($Mbps$) and a maximum of 150 $Mbps$. The median, average, and standard deviation are 108.5, 105.385, and 29.79, respectively. The memory dirtying rate $(d_{i,j})$ is another parameter for which we consider non-average values. The minimum value is 0.02323 $Mbps$, and the maximum is 145.076 $Mbps$. The median and average memory dirtying rates are 18.52 and 28.979, respectively, with a standard deviation of 31.89. Memory sizes for VMs and containers $(M_j)$ range from a minimum of 249.41 $megabytes$ ($MB$) to a maximum of 4080.94 $MB$. The median, average, and standard deviation are 813.326, 1049.83, and 625.268, respectively. The final parameter considered is $\lambda_{i,j}$, which is the memory dirtying rate divided by the transfer rate. During the migration process, the minimum and maximum $\lambda_{i,j}$ are 0.000196718 and 0.999938322. The median and average are 0.166656325 and 0.274938801, respectively, with a standard deviation of 0.292347702.

### 5.1 Results using average parameter values and the dataset

We investigate the performance of the pre-copy and MiGrror using average parameter values of VM/container size, memory dirtying rate, and transfer rate for each VM/container in this subsection. Figure 2 illustrates that downtime is the most noticeable distinction between the pre-copy and MiGrror. In our experiments, the median downtime for pre-copy is 265.924 $ms$, while the median downtime for MiGrror is less than 1 $ms$. The pre-copy downtime is unacceptable for future 5G and 6G delay-sensitive applications since it results in prolonged service interruptions during migration. The MiGrror technique, on the other hand, generates much less downtime than the pre-copy technique since it uses live mirroring between the source and destination.

Furthermore, as illustrated in the figure, the migration time using the MiGrror technique is less than that of the pre-copy technique, with maximums of 23.65 $seconds$ ($s$) and 23.93 $s$, respectively. Using non-average parameter values in the subsequent subsection reveals a significantly larger difference. With the shorter migration time, resources at the source can be made available to other containers more quickly.

Although the MiGrror reduces downtime and migration time, it comes at a cost: migration overhead. However, the cost is negligible. Since the MiGrror method synchronizes faster than the pre-copy method, it requires more bandwidth to mirror changes from the source to the destination. The MiGrror consumes more bandwidth than pre-copy, but it is only a negligible 1.16% increase in total migration overhead. Despite the additional overhead of 1.16%, downtime and migration time are reduced.

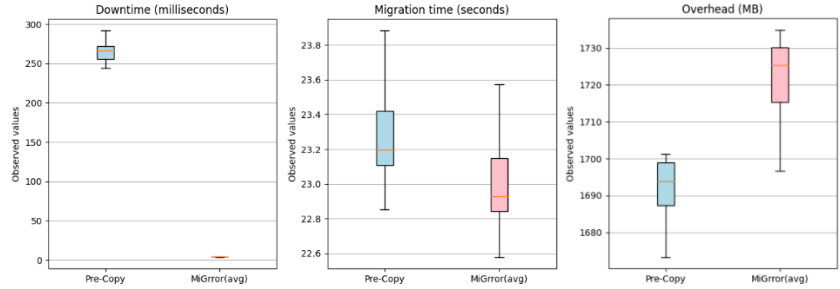

Fig 2.  Comparison of the pre-copy and MiGrror migration methods using the dataset's average parameter values (blue: Pre-Copy, red: MiGrror) (left: Downtime, middle: Migration Time, right: Migration Overhead)

## 5.2  Non-average parameter value results using the dataset

In this subsection, we examine the performance of the pre-copy and MiGrror techniques. For the MiGrror method, we use average and non-average parameter values for memory dirtying rate, transfer rate, and VM/container size and compare them to the results of the classical pre-copy model, which only uses average parameter values.

We cannot compare non-average parameter values of MiGrror and pre-copy since the pre-copy uses a limited number of rounds, and the stated parameters can change many times during each round. Therefore, only non-average parameter values of the MiGrror migration method, and average parameter values of both the pre-copy and MiGrror migration methods, are presented in Figure 3.

Figure 3 is the best representation of why we must consider non-average parameter values in contrast to the traditional view of using only average parameter values. The figure depicts that fluctuations of downtime, migration time, and migration overhead are unanticipated even when using the same method, MiGrror. To clarify, when using average and non-average parameter values, neither the mean nor the median of the results is identical or even close. The same pattern holds true for the maximum and the minimum of the results.

Consider the MiGrror results with average and non-average parameter values. Although the median downtime for average parameter values is roughly four times that of non-average parameter values, the maximum downtime using average parameter values is roughly 50% of non-average parameter values. Furthermore, the maximum migration time using average parameter values and non-average parameter values differs by about 25%, while the minimum migration time using average and non-average parameter values differs by more than 32%. Moreover, when using average parameter values, the standard deviation of the downtime is only 0.21, whereas when using non-average parameter values, it is 3.81. This high standard deviation results from memory dirtying rate of the last event, which directly affects downtime, according to the model and results. The standard deviation of migration time follows the same pattern, when using average and non-average parameter values at 0.30 and 3.46, respectively. The difference in standard deviation between average and non-average parameter values for migration overhead is substantial; 22.36 and 346.33, respectively. Consequently, using non-average parameter values provides us with a new perspective to improve future applications and prevent unanticipated outcomes, such as those shown in Figure 3, when implementing them in the real world. These findings highlight the practicality of using non-average parameter values when analyzing data.

We demonstrate that the MiGrror migration overhead is only 0.5% greater than the pre-copy migration overhead on average, which is an additional advantage of utilizing non-average parameter values. However, when using average parameter values, the MiGrror overhead amount is 1.19% greater than the pre-copy migration overhead on average.

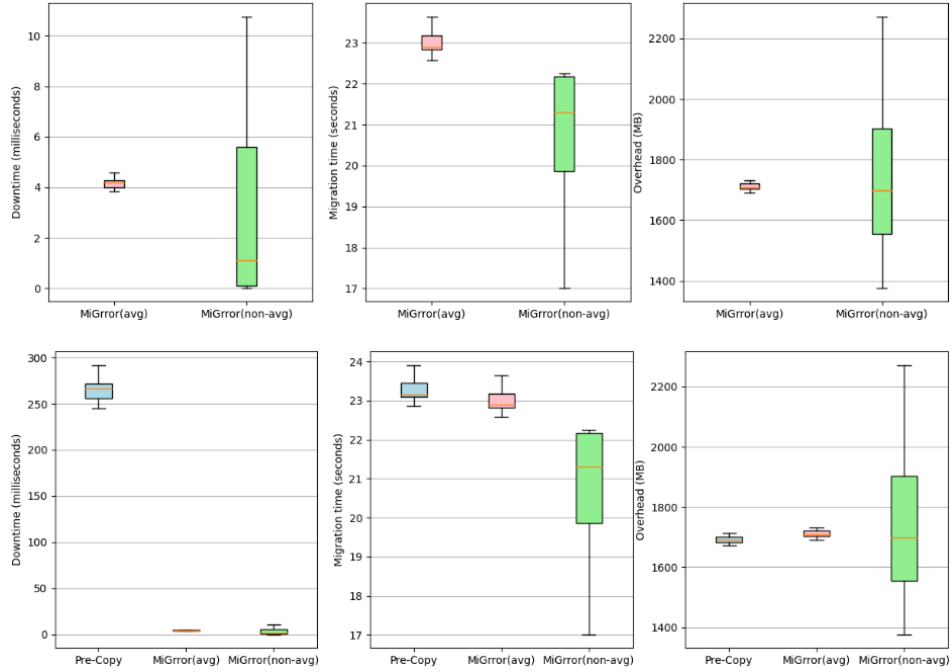

**Fig 3.** Comparison of the dataset's average and non-average parameter values using MiGrror migration (left: Downtime, middle: Migration Time, right: Migration Overhead, red: MiGrror using average values, green: MiGrror using non-average values)

## 5.3 Non-dataset parameter values

This section studies the performance of the pre-copy and MiGrror models in terms of several parameters: container size, memory dirtying rate, and transfer rate. We use average parameter values in this subsection to make a fair comparison since we cannot calculate the pre-copy results using the non-average parameter values as explained in the previous subsection. Each subsection focuses on one of these three parameters, with the final subsection focusing on $\lambda_{i,j}$ variations.

### 5.3.1 Performance based on varying VM/container size

Figure 4 illustrates the impact of varying VM/container sizes on downtime, migration time, and migration overhead. The first row of 3-D figures represents the variations in VM/container size in terms of different $\lambda_{i,j}$ ratios and the corresponding results. The figure also depicts the downtime, migration time, and migration overhead for various VM/container sizes with $\lambda_{i,j}$ set to 0.25 in the middle row and 0.5 in the bottom row.

This figure shows that **increasing the VM/container** size has **no meaningful** effect on MiGrror **downtime**. Only with high $\lambda_{i,j}$ ratios does pre-copy downtime increase, but the increase is negligible since the difference is less than 4% between the smallest and largest VM/container sizes. However, increasing the VM/container size increases migration time and overhead. It is also evident that the difference between migration time and overhead of the researched methods becomes more apparent with a higher $\lambda_{i,j}$. In addition, as shown in the figure, migration time and overhead increase as the VM/container size increases.

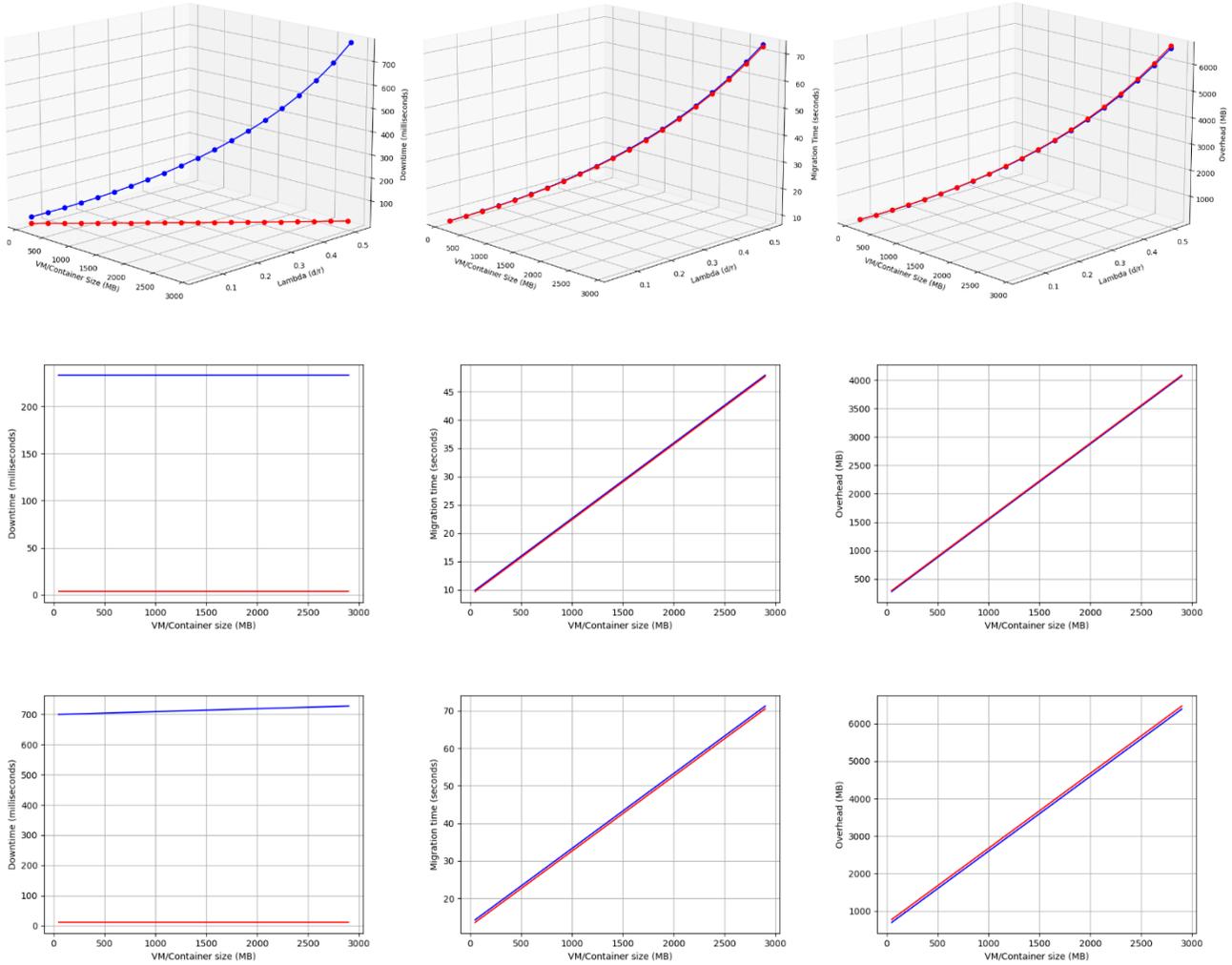

**Fig 4.** Downtime, Migration Time, and Migration Overhead as functions of VM/container size for the pre-copy and MiGrror migration methods using non-dataset parameter values (blue: Pre-Copy, red: MiGrror) (top: $\lambda_{i,j}$ varies between 0.04 and 0.515, middle: $\lambda_{i,j}$ is constant and set to 0.25, bottom: $\lambda_{i,j}$ is constant and set to 0.5)

### 5.3.2 Performance based on variation of transfer rate

Figure 5 illustrates how various transfer rates affect downtime, migration time, and migration overhead for the pre-copy and MiGrror migration methods. We consider 200 $MB$ as a typical size for a container which also is considered a lightweight VM. Furthermore, we consider that there is a 1000 $Mbps$ total bandwidth for all VMs/containers. The VM/container size and memory dirtying rate are fixed at 200 $MB$ and 50 $Mbps$ (since we assume there are 20 VMs/containers), respectively. The figure represents that increasing transfer rates reduces pre-copy downtime significantly, from 1421 $milliseconds$ ($ms$) to 70 $ms$, but the value is still relatively high. However, MiGrror downtime begins at 22 $ms$ and ends at 1.1 $ms$, which is still superior to the pre-copy. In terms of migration time, the MiGrror is 5.65% less than that of pre-copy at low transfer rates, and less than 1% at high transfer rates. In terms of migration overhead, the MiGrror consumes 14.61% more bandwidth than pre-copy at a low transfer rate, but this decreases to 4.15% more bandwidth consumption at a high transfer rate.

These findings imply that pre-copy downtime is still unacceptable for latency-sensitive mobile IoT applications, even with a high transfer rate. In addition, increasing the transfer rate converges the migration time and overhead of both methods. These results indicate that when the transfer rate is high, and the downtime is nonessential, such as in applications with no delay constraints. The performance of both approaches is nearly identical. However, when the transfer rate is limited, MiGrror outperforms pre-copy in terms of downtime and migration time.

Furthermore, as shown in the figure, migration time and overhead decrease as the transfer rate rises. The same pattern applies to pre-copy downtime but not MiGrror downtime. MiGrror downtime is low from the beginning.

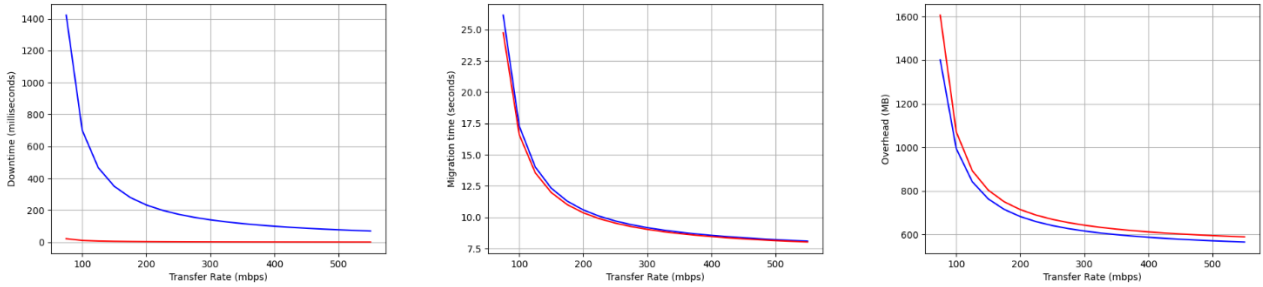

**Fig 5.** Downtime, Migration Time, and Migration Overhead as functions of Transfer Rate for the pre-copy and MiGrror migration methods using non-dataset parameter values (blue: Pre-Copy, red: MiGrror)

### 5.3.3 Performance based on variation of memory dirtying rate

Figure 6 illustrates how various memory dirtying rates affect downtime, migration time, and migration overhead for the pre-copy and MiGrror migration methods. The VM/container size and transfer rate are fixed at 200 $MB$ and 200 $Mbps$, respectively. The figure depicts that increasing memory dirtying rates increases pre-copy downtime significantly, from around 20 $ms$ at a 5 $Mbps$ transfer rate to more than 700 $ms$ at a 100 $Mbps$ transfer rate. However, MiGrror downtime begins at less than 1 $ms$ and ends at 11 $ms$ with the same transfer rates, which is still superior to the pre-copy. In terms of migration time, the MiGrror is 0.21% less than that of the pre-copy at a low transfer rate and the migration time improves to 4.72% at a high transfer rate. In terms of migration overhead, the MiGrror consumes 0.91% more bandwidth than pre-copy at a low transfer rate, but this increases to about 9% more bandwidth consumption at a high transfer rate.

These findings imply that pre-copy downtime is still unacceptable for latency-sensitive mobile IoT applications, even with a low memory dirtying rate. In addition, decreasing the transfer rate converges migration time and overhead of both methods. These results indicate that when the memory dirtying rate is low and the downtime is nonessential, such as in applications with no delay constraints, the performance of both approaches is nearly identical. However, when the memory dirtying rate is high, MiGrror outperforms pre-copy in terms of downtime and migration time.

Furthermore, as shown in the figure, migration time and overhead increase as the memory dirtying rate rises. The same pattern applies to pre-copy downtime but not MiGrror downtime.

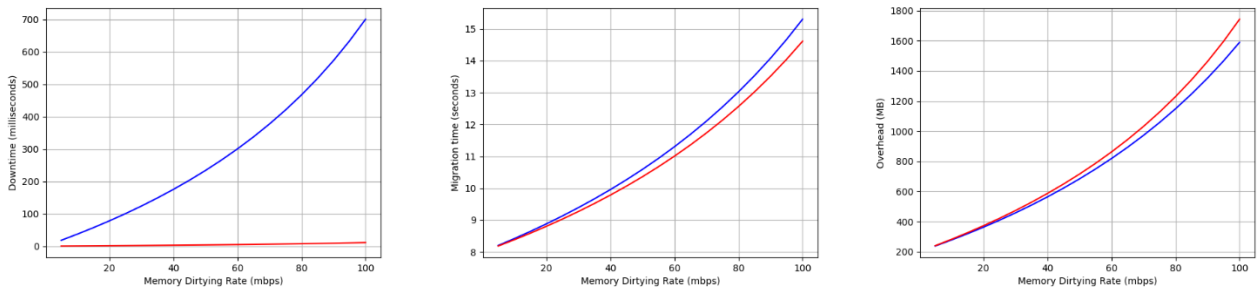

**Fig 6.** Downtime, Migration Time, and Migration Overhead as functions of Memory Dirtying Rate for the pre-copy and MiGrror migration methods using non-dataset parameter values (blue: Pre-Copy, red: MiGrror)

### 5.3.4 Performance based on variation of $\lambda_{i,j}$ (memory dirtying rate divided by transfer rate)

Figure 7 illustrates how various $\lambda_{i,j}$ rates affect downtime, migration time, and migration overhead for the pre-copy and MiGrror migration methods. The VM/container size is fixed at 200 $MB$. The figure shows increasing $\lambda_{i,j}$ rates increase pre-copy downtime significantly, from 60 $ms$ at a 0.08 $\lambda_{i,j}$ rate to more than 700 $ms$ at a 0.50 $\lambda_{i,j}$ rate.

However, MiGrror downtime begins at 0.95 $ms$ and ends at 11 $ms$ with the same $\lambda_{i,j}$ rates, which is still superior to the pre-copy. In terms of migration time, the MiGrror is 0.69% less than that of the pre-copy at a low $\lambda_{i,j}$ rate and the difference is raised to more than 6.65% at 0.65 $\lambda_{i,j}$ rate. In terms of migration overhead, the MiGrror consumes 0.98% more bandwidth than the pre-copy at a 0.0275 $\lambda_{i,j}$ rate, but this increases to a 9.63% more bandwidth consumption at a 0.5 $\lambda_{i,j}$ rate.

These findings imply that pre-copy downtime is still unacceptable for latency-sensitive mobile IoT applications, even with a low $\lambda_{i,j}$ rate. In addition, decreasing the transfer rate converges migration time and overhead of both methods. These results indicate that when the $\lambda_{i,j}$ rate is low, and the downtime is nonessential, such as in applications with no delay constraints, the performance of both approaches is nearly identical. However, when the $\lambda_{i,j}$ rate is high, MiGrror outperforms pre-copy in terms of downtime and migration time. In fact, only when the $\lambda_{i,j}$ is low the pre-copy downtime is acceptable.

Furthermore, as shown in the figure, migration time and overhead increase as the $\lambda_{i,j}$ rate rises. The same pattern applies to pre-copy downtime but not MiGrror downtime.

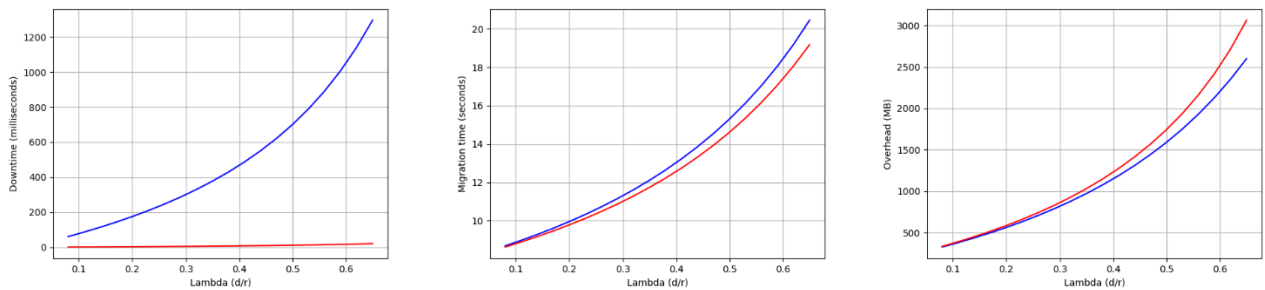

**Fig 7.** Downtime, Migration Time, and Migration Overhead as functions of $\lambda_{i,j}$ for the pre-copy and MiGrror migration methods using non-dataset parameter values (blue: Pre-Copy, red: MiGrror)

5.3.5 Further Discussions

Figure 8 shows the relationship between downtime and migration duration for the pre-copy and MiGrror migration methods. As pre-copy downtime increases, migration time increases linearly. This finding means that we cannot use the pre-copy method after a certain value when application downtime or migration time is critical. For instance, if an application cannot tolerate more than 100 $ms$ without interrupting users, we cannot use the pre-copy for that application, even if the migration time falls within that application's tolerance range. Using MiGrror, however, the rate of increase in downtime is significantly lower than its migration time. This finding indicates that by employing the MiGrror method, the MEC is able to service applications, as in the stated example, with a greater amount of migration time, since the MiGrror downtime is still within the tolerance range of the application.

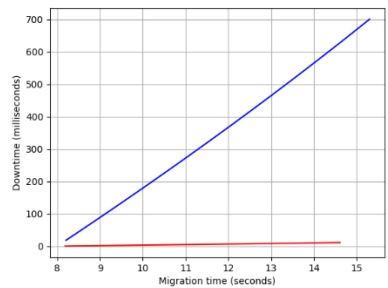

**Fig 8.** The pre-copy and MiGrror Downtime as a function of their Migration Time (blue: Pre-Copy, red: MiGrror)

Furthermore, as shown in figures 5, 6, and 7 as well as in the modelling results, the memory dirtying rate directly impacts the amount of memory transfer at each MiGrror migration event. The MiGrror method reduced downtime by lowering the final amount of memory transfer, as shown in the results and equation (23).

Moreover, based on equations (7, 9, 23, 25) and figures 5 and 7, it is evident that with the same overall bandwidth, increasing the number of containers will reduce the transfer rate of each container, resulting in longer migration times. The transfer rate also slightly increases container downtimes. We skip the figure for the preceding argument since it can be inferred from figures 5 and 7.

# 6 Conclusion and future directions

In this paper, we model the MiGrror method for multi-service migrations for the first time. We use non-average parameter values as well as traditional average parameter values for downtime, migration time, and migration overhead for the first time. We illustrated that the MiGrror migration time and downtime outperform the pre-copy ones. As demonstrated in the paper, utilizing non-average parameters allows for a better understanding of what occurs during migration and more accurate results. Outputs can deviate drastically during crucial migration phases when actual (non-average) input parameters vary while their averages are unchanged. Using non-average input parameter values in migration models can provide a refined, rational, migration analysis, which takes into account low migration time, downtime, and mobility, in multi-containerized edge computing environments. As a result, we use non-average parameter values to obtain more accurate results and comprehension of the environment.

Furthermore, the proposed migration model can be integrated with other migration approaches with improved accuracy, such as compression. In addition, the new migration model can be integrated with ML techniques to provide a detailed view of expected outputs. As a result of utilizing MiGrror and the proposed strategies, we showed that MiGrror improves service continuity and availability for users. This study will lead to future research exploring models using non-average parameter values in order to better understand and optimize downtime and migration time using the MiGrror method.

## Acknowledgments

The Natural Sciences and Engineering Research Council of Canada and Bell Canada partially fund this work.